\documentclass[usenatbib]{mn2e}
\usepackage{color}
\usepackage{graphicx}
\usepackage{amssymb,amsmath}
\usepackage{natbib}

\title[Diffuse Ly$\alpha$ Halos at $z=2.2-6.6$]
 {
Diffuse Ly$\alpha$ Halos around Galaxies at $z=2.2-6.6$: Implications for Galaxy Formation and Cosmic Reionization
}

\author[R. Momose et al.]
{
Rieko~Momose$^1$,
Masami~Ouchi$^{1,2}$,
Kimihiko~Nakajima$^{2,3}$,
Yoshiaki~Ono$^1$,
\newauthor
Takatoshi~Shibuya$^{1,4}$,
Kazuhiro~Shimasaku$^{3,5}$, 
Suraphong~Yuma$^1$,
\newauthor
Masao~Mori$^4$,
and
Masayuki~Umemura$^4$
\newauthor
\\
$^1$
Institute for Cosmic Ray Research, The University of Tokyo, 
5-1-5 Kashiwanoha, Kashiwa, Chiba 277-8583, Japan \\
$^2$
Kavli Institute for the Physics and Mathematics of the Universe (WPI), The University of Tokyo, 
5-1-5 Kashiwanoha, Kashiwa, \\
Chiba 277-8583, Japan \\
$^3$
Department of Astronomy, Graduate School of Science, The University of Tokyo, 
7-3-1 Hongo, Bunkyo-ku, Tokyo 113-0033, Japan \\
$^4$
Center for Computational Sciences, The University of Tsukuba, 
1-1-1 Tennodai, Tsukuba, Ibaraki 305-8577, Japan \\
$^5$
Research Center for the Early Universe, Graduate School of Science, 
The University of Tokyo, Tokyo 113-0033, Japan \\
}

\date{}

\pagerange{\pageref{firstpage}--\pageref{lastpage}} \pubyear{2014}

\begin{document}

\label{firstpage}

\maketitle

\begin{abstract}
We present diffuse Ly$\alpha$ halos (LAHs) identified in the composite Subaru narrowband images of $100-3600 $ Ly$\alpha$ emitters (LAEs) at $z=2.2$, $3.1$, $3.7$, $5.7$, and $6.6$. First, we carefully examine potential artifacts mimicking LAHs that include a large-scale point-spread function (PSF) made by instrumental and atmospheric effects. Based on our critical test with composite images of non-LAE samples whose narrowband-magnitude and source-size distributions are the same as our LAE samples, we confirm that no artifacts can produce a diffuse extended feature similar to our LAHs.
After this test, we measure the scale lengths of exponential profile 
for the LAHs estimated from our $z=2.2-6.6$ LAE samples of 
$L_{\rm Ly\alpha}\gtrsim 2\times 10^{42}$ erg s$^{-1}$.
We obtain the scale lengths of $\simeq 5-10$ kpc at $z=2.2-5.7$,
and find no evolution of scale lengths in this redshift range beyond our measurement uncertainties.
Combining this result and the previously-known UV-continuum size evolution, we infer that the ratio of LAH to UV-continuum sizes is nearly constant at $z=2.2-5.7$. 
The scale length of our $z=6.6$ LAH is larger than $5-10$ kpc just beyond the error bar, which is a hint that the scale lengths of LAHs would increase from $z=5.7$ to $6.6$. If this increase is confirmed by future large surveys with significant improvements of statistical and systematical errors, this scale length change at $z\gtrsim 6$ would be a signature of increasing fraction of neutral hydrogen scattering Ly$\alpha$ photons, due to cosmic reionization.
\end{abstract}


\begin{table*}
\centering
\begin{minipage}{168mm}
\begin{center}
\caption{Samples and Diffuse Ly$\alpha$ Halos in Our and Previous Studies}
\label{tab:z}
\begin{tabular}{cccccccccc}
\hline
Redshift	 & $N$ 	& SB limit & $C_n$ & $r_n$	 & Reference	\\
(1) 		 & (2) 	& (3) 	   & (4) 	     & (5)	 	 &  (6)	\\
\hline
2.2		& 3556$^\dagger$ 	& 
0.16
	& 1.5	& 7.9$^{+0.56}_{-0.49}$ 	& This study	\\
3.1 		& 316	& 
1.7
	& 5.3	& 9.3$^{+0.48}_{-0.53}$ 	& This study	\\
3.7 		& 100	& 
2.8
	& 
---
	& 
---
& This study	\\
5.7 & 397	& 
0.55
	& 2.0	& 5.9$^{+0.65}_{-0.53}$ 		& This study	\\
6.6 & 119	& 
1.8
	& 0.8	& 12.6$^{+3.3}_{-2.4}$ 		& This study \\
\hline
2.06 & 187	& $\sim$10 & $4-15$	& $3.7-5.7$			& \citet{feld13}$^a$ \\
2.65 & 92	& $\sim$1& 2.5 & 25.2							& \citet{stei11}$^b$	\\
3.1	& 22	& --		& --		& --								& \citet{haya04}$^c$ \\	
3.1	& 130	& -- 	& 0.7 	& 20.4					& \citet{matsu12} $^d$\\  
3.1	& 237	& -- 	& 1.4 	& 13.2						& \citet{matsu12} $^d$\\  
3.1	& 861	& -- 	& 1.4 	& 10.7						& \citet{matsu12}$^d$ \\  
3.1	& 864	& -- 	& 1.5 	& 9.1						& \citet{matsu12}$^d$ \\  
3.10 & 241	& $\sim$7 & $15-38$	& $5.5-6.0$			& \citet{feld13}$^e$ \\
3.21 & 179	& $\sim$7 & $12-31$	& $2.8-8.4$			& \citet{feld13}$^e$ \\
\hline
\end{tabular}\\
\end{center}
(1) Redshift; 
(2) the number of LAEs used for stacking analyses; 
(3) one sigma SB limits for mean-combined images in units of $10^{-19}$ erg s$^{-1}$ cm$^{-2}$ arcsec$^{-2}$;
(4) best-fit $C_n$ for mean-combined images in units of $10^{-18}$ erg s$^{-1}$ cm$^{-2}$ arcsec$^{-2}$;
(5) best-fit $r_n$ for mean-combined images in units of kpc;
(6) reference for previous study results.
All of the values in (3)-(5) are measured by the mean-combined method, except for those in \citet{matsu12} who only use the median-combined method for their images down to the one sigma SB limits of 
$\sim 0.3-2.0 \times 10^{-19}$ erg s$^{-1}$ cm$^{-2}$ arcsec$^{-2}$.
The best-fit $r_n$ values of our LAEs from the median-combined images
are 
11.1$^{+1.2}_{-0.97}$, 6.3$^{+2.5}_{-1.4}$,
7.7$^{+1.9}_{-1.3}$ , and 13.9$^{+0.75}_{-0.99}$ kpc 
at $z=2.2$, $3.1$,
$5.7$, and 6.6, respectively.
$^\dagger$: Our $z=2.2$ sample consists of 3556 LAEs. For the LAH evolution discussion, 
the values of (3)-(5) in this line correspond to
those of 2115 LAE subsample with the Ly$\alpha$ luminosity limit of $1\times 10^{42}$ erg s$^{-1}$.
 The SB limit, $C_n$, and $r_n$ values for our 3556 LAEs are
$0.13\times 10^{-19}$ erg s$^{-1}$ cm$^{-2}$ arcsec$^{-2}$,
$0.8 \times 10^{-18}$ erg s$^{-1}$ cm$^{-2}$ arcsec$^{-2}$,
and 10.0$^{+0.38}_{-0.36}$ kpc, respectively.
$^a$: \citet{feld13} claim that their analysis finds no evidence of LAH at $z=2.06$.
$^b$: \citet{stei11} also estimate the scale length of their LAH by the median-combined method to be 17.5 kpc.
$^c$: Scale lengths of LAH are not measured in \citet{haya04} .
$^d$: The four samples of LAEs in the highest to the lowest density environments
are shown from top to the bottom lines.
$^e$: \citet{feld13} argue that their LAHs at $z\sim3.1$ are marginally detected. \\
\end{minipage}
\end{table*}

\section{Introduction}
\label{introduction}
The circum-galactic medium (CGM)  is closely related to galaxy formation and evolution. Gas inflows into galaxies could trigger starbursts (e.g., \citealp{dek09a,dek09b}), while gaseous outflows are thought to be a physical process of quenching star formation (e.g., \citealp{mori02,scan05,mori06,dave11}). The distribution of the CGM is characterized by Ly$\alpha$ emission, because Ly$\alpha$ photons escaping from a galaxy are resonantly scattered by surrounding neutral hydrogen gas. The scattered light would produce diffuse Ly$\alpha$ emission around a galaxy. This spatially-extended Ly$\alpha$ emission is dubbed Ly$\alpha$ halo (LAH). Numerical simulations have predicted that LAHs are ubiquitously present around high-$z$ galaxies (e.g., \citealp{lau07,zhen11,dijk12,verha12}).

Radial surface brightness (SB) profiles of the LAHs are useful to understand kinematic properties of CGM and neutral hydrogen fraction of inter-galactic medium (IGM) at the epoch of cosmic reionization. \citet{zhen11} have predicted that a slope of radial SB profile depends on an outflowing velocity of CGM based on their radiative transfer model. \citet{dijk12} have calculated radiative transfer of Ly$\alpha$ photons propagating through clumpy and dusty large-scale outflowing inter-stellar medium (ISM), and reproduced an extended Ly$\alpha$ structure. Furthermore, \citet{jee12} demonstrate that radial SB profiles of LAHs are flatter at the epoch of reionization than at the post reionization epoch, due to Ly$\alpha$ photons scattered by neutral hydrogen of IGM. 
On the other hand, recent observations find no diffuse metal-line emission of hot ionized gas around
high-$z$ galaxies on average \citep{yuma13}, and indicate that LAHs are probably not 
made by emission of hot CGM given by outflow, but the other physical processes of cold CGM.

Extended Ly$\alpha$ emission has been observed around nearby star-forming galaxies (e.g., \citealp{ost09,hay13a,hay13b}) and QSOs (e.g., \citealp{rau08,goto09}). However, LAHs are too diffuse and faint to be detected for high-$z$ galaxies on an individual basis. LAHs at $z \geq 2$ are found in stacked data of $\sim20-2000$ narrowband (NB) images of high-$z$ galaxies in previous studies. \citet{haya04} have discovered a LAH around Lyman break galaxies (LBGs) at $z = 3.1$ with their composite NB image.
\citet{stei11} have identified extended LAHs with a radius of $r \sim 80$ kpc around LBGs at a spectroscopic redshift of $\langle z\rangle=2.65$ by stacking $92$ NB images. \citet{matsu12} have stacked 
130-864 LAEs at $z=3.1$, and detected LAHs. On the other hand, \citet{jian13} have found no extended Ly$\alpha$ emission
in their composite image produced with dozens of LAEs at $z=5.7$ and $6.6$, 
although their results are based on the small statistics.

There is an argument of systematic uncertainties producing
spurious features similar to LAHs \citep{feld13}. 
\citet{feld13} have claimed that one of major sources of spurious LAHs is a large-scale point-spread function (PSF) that appears in deep images taken by ground-based observations \citep{king71}. A profile of large-scale PSF is largely extended, and the slope of profile changes at large radii of $>4\arcsec$ \citep{feld13}, probably due to atmospheric turbulence and instrumental conditions (e.g., \citealp{raci96,bern07}). The profile of large-scale PSF can mimic that of LAH, and would be mistakenly identified as an LAH. Thus, the existence of LAHs is still under debate.
In order to test the existence of LAHs, a careful data analysis as well as a large galaxy sample is required.

Here, we present our analysis and results of LAHs at $z=2.2-6.6$ based on our large LAE samples given by Subaru NB observations (\citealp{ouchi08,ouchi10,naka12}). The large LAE samples of high-quality Subaru images enable us to test the existence of diffuse LAHs and to extend the study from $z\sim 2-3$ to $6.6$.
We show the data and analysis in Section \ref{data_and_analysis}, systematic errors in Section \ref{systematic_errors}, and
our results of LAHs in Section \ref{results}, and discuss 
galaxy formation and reionization in Section \ref{discussions}.
We summarize our results and discussions in Section \ref{summary}.
Throughout this paper, we use AB magnitudes and 
adopt a cosmology parameter set of
($\Omega_m$, $\Omega_\Lambda$, $H_0$) = ($0.3$, $0.7$, $70$ km s$^{-1}$ Mpc$^{-1}$).
 In this cosmology, $1\arcsec$ corresponds to transverse sizes of
 ($8.3$, $7.6$, $7.2$, $5.9$, $5.4$) kpc at $z=$ ($2.2$, $3.1$, $3.7$, $5.7$, $6.6$).
\begin{figure*} 
	\begin{center}
		\includegraphics[scale=0.25]{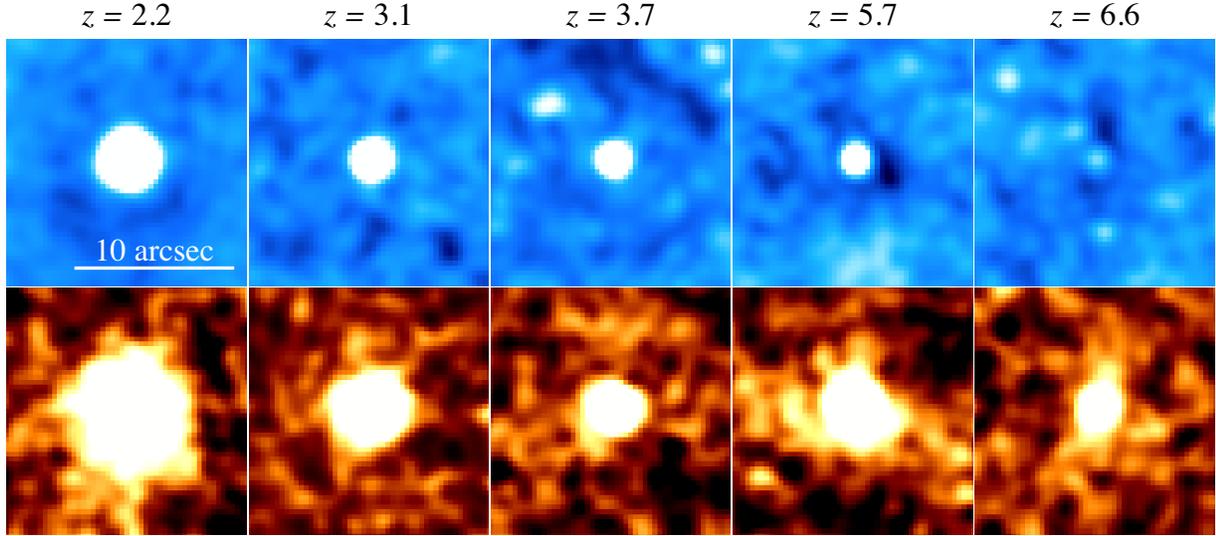}
	\end{center}
	\caption{
	Composite continuum (top panels) and Ly$\alpha$ (bottom panels) images of our LAEs
	produced by the mean-combined method.
	From left to right panels, we show $z=2.2$, $3.1$, $3.7$, $5.7$, and $6.6$ LAE images.
	}
	\label{fig:image_mean}
\end{figure*}

\begin{figure*}
	\begin{center} 
		\includegraphics[scale=0.25]{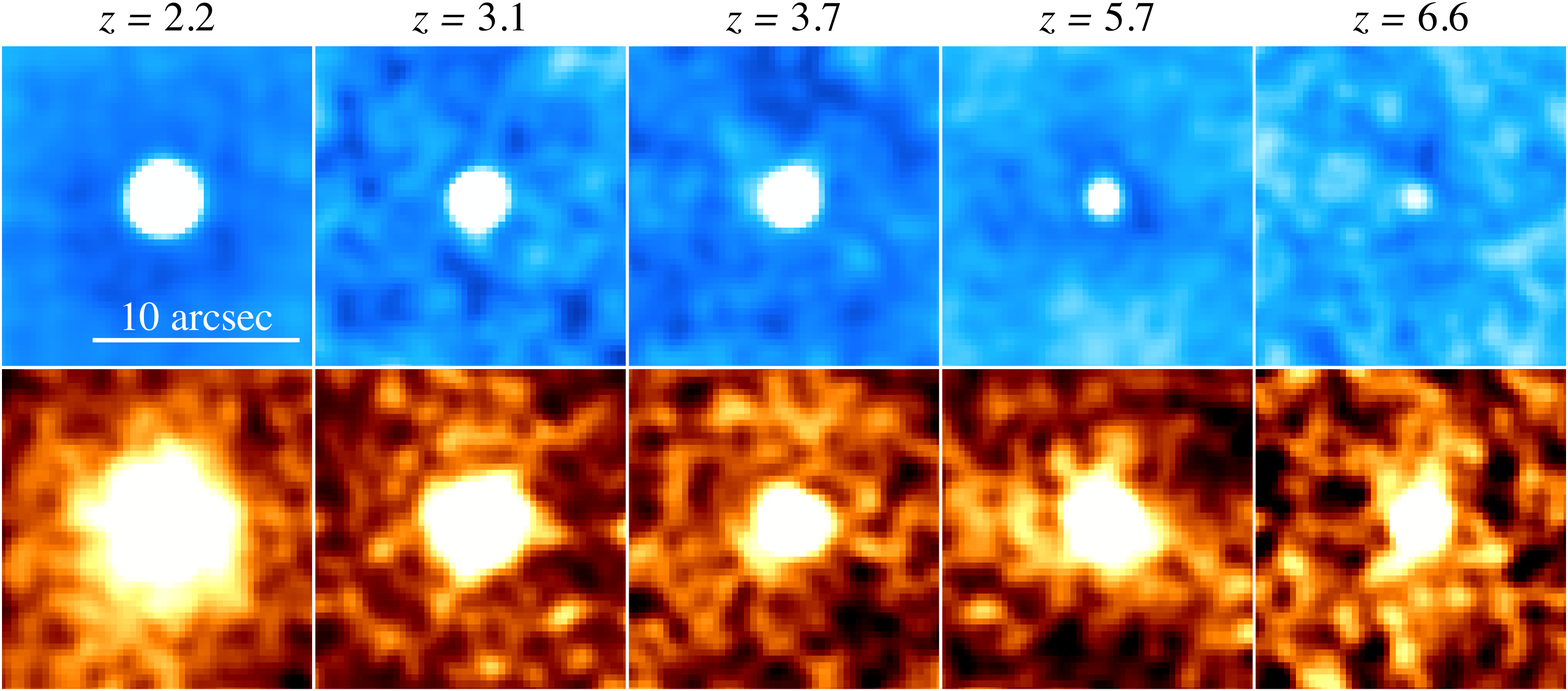}
	\end{center}
	\caption{
	Same as Figure \ref{fig:image_mean}, but for the median-combined method.
	}
	\label{fig:image_median}
\end{figure*}

\begin{figure*}
	\begin{center} 
		\includegraphics[scale=0.35]{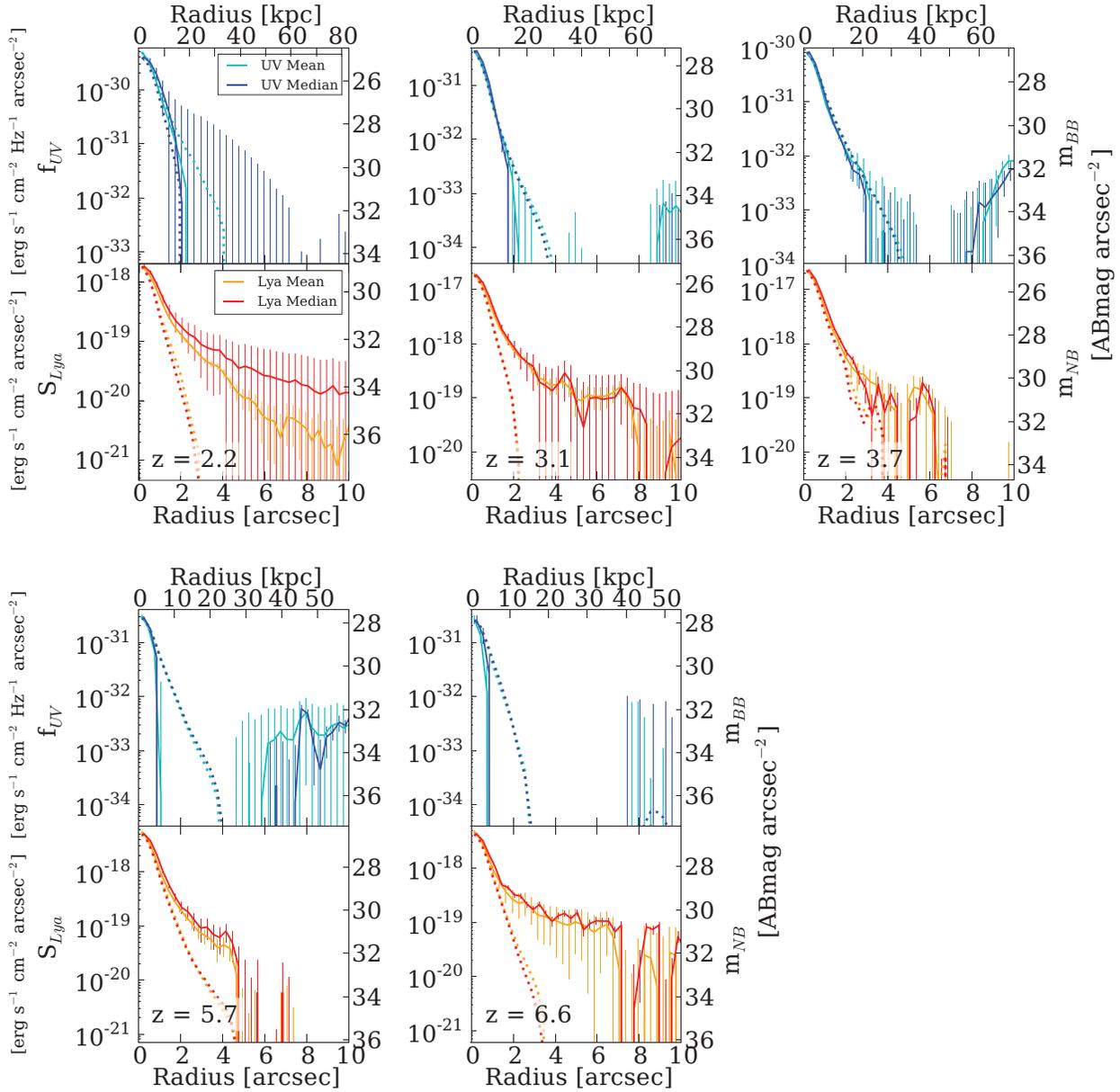}
	\end{center}
	\caption{
	Radial SB profiles of composite images of LAEs (solid lines) and PSFs (dotted lines) at redshifts of $z=2.2-6.6$. 
	The upper and lower panels represent SB 
	profiles of continuum and Ly$\alpha$ emission, respectively. The cyan and orange (blue and red) lines denote the results of mean-combined 
	(median-combined) methods.
	}
	\label{fig:radi_allz}
\end{figure*}

\section{Data and Analysis}
\label{data_and_analysis}
\subsection{Data Set}
We use large photometric samples of LAEs at $z=2.2$, $3.1$, $3.7$, $5.7$, and $6.6$ made by the large-area NB imaging surveys of Subaru telescope.
Our $z=2.2$ sample consists of $3556$ LAEs found in five deep fields of COSMOS, GOODS-N, GOODS-S, SSA22, and SXDS \citep{naka12}. The total area of the deep fields with our $z=2.2$ LAEs is about $2.3$ deg$^2$. The $z=2.2$ LAEs are identified by an excess of flux in an NB of $NB387$ whose central wavelength and FWHM are $3870$ {\AA} and $94$ {\AA}, respectively. 
The continua of these LAEs are determined with $V$-band images taken by \citet{cap04}, \citet{haya04}, \citet{tani07}, \citet{furu08}, and \citet{tayl09}.
Our $z=3.1-6.6$ LAE samples are obtained only in the 1 deg$^2$ SXDS field \citep{ouchi08,ouchi10}. There are ($316$, $100$, $397$, $119$) LAEs at $z=$ ($3.1$, $3.7$, $5.7$, $6.6$) identified with NBs of ($NB503$, $NB570$, $NB816$, $NB921$). The central wavelength and FWHM values are ($5029$\AA, $74$\AA), ($5703$\AA, $69$\AA), ($8150$\AA, $120$\AA), and ($9196$\AA, $132$\AA) for $NB503$, $NB570$, $NB816$, and $NB921$, respectively. The continua of LAEs are estimated with 
broadband images of $R$, $i'$, $z'$, and $J$ bands for $NB503$, $NB570$, $NB816$, and $NB921$ LAEs. We refer to these broadband images for our continuum estimates as continuum images.
These optical and near-infrared images are taken from the public data of SXDS (\citealp{furu08}) and UKIDSS (\citealp{law07}), respectively.
All of the imaging data used in this study are obtained with Subaru/Suprime-Cam, except for the $J$-band image.
The $J$-band observations are conducted with the the Wide Field Camera (WFCAM; \citealp{hew06,cas07}) on the UK Infrared Telescope (UKIRT). 
In summary, our samples have a total of 4488 LAEs at $z=2.2-6.6$ on the $2.3$ deg$^2$ sky.
Our LAE samples have the Ly$\alpha$ luminosity and equivalent-width limits
of $\sim 10^{42}$ erg s$^{-1}$ and $\sim 20-60$\AA, respectively (Section \ref{scale_lejgths_of_our_lahs_and_comparisons}; 
see \citealt{ouchi08,ouchi10,naka12} for more details).
Our LAE samples include 83, 41, 26, 17, and 16 spectroscopically-confirmed LAEs at $z=2.2$, 
$3.1$, $3.7$, $5.7$, and $6.6$, respectively (\citealp{ouchi08,ouchi10,naka12,naka13,hashimoto13,shib14b}; Nakajima et al. 2014 in prep.).
These spectroscopic studies find that the contamination rate of these LAE samples is negligibly small,
$\sim 0$\%. The contamination rate can be up to only $\sim 30$\%, even if 
all of the unidentified objects near the flux limits of spectroscopy 
are regarded as contamination sources.
No redshift dependence of contamination rate is reported in these spectroscopic studies.
Because our LAE samples should include some contamination sources,
the effect of contamination is discussed with the results of mean and median statistics
in Section \ref{scale_lejgths_of_our_lahs_and_comparisons}.

\subsection{Image Stacking}
\label{image_stacking}
To investigate LAHs, we carry out stacking analysis with the continuum and NB images of our LAEs.

\begin{description}
\item[(i) Smoothing Images]\mbox{}\\ 
We smooth all the continuum and NB images with Gaussian kernels to match their seeing sizes to an FWHM of $1\farcs32$ 
that is the largest PSF size among the images. 
\vspace{0.5em}
\item[(ii) Subtracting Continuum Fluxes]\mbox{}\\ 
We subtract the smoothed continuum images from the smoothed NB images under the assumption of $f_\nu = $const. The NB images with a continuum subtraction is referred to as Ly$\alpha$ images.
\vspace{0.5em}
\item[(iii) Making Cutout Images]\mbox{}\\ 
We make cutout images of the smoothed continuum and Ly$\alpha$ images with a size of $45\arcsec\times45\arcsec$ centered 
at a position of LAE. Here, we exclude LAEs that are placed at the areas within 200 pixels from the edge of imaging data
to avoid systematic effects. 
The numbers of our LAEs that we use are listed in Table \ref{tab:z}.
\vspace{0.5em}
\item[(iv) Stacking Images]\mbox{}\\ 
We stack these cutout images with the {\tt imcombine} task of {\tt IRAF} in two ways, ``mean-combined'' and ``median-combined'' methods. In the mean-combined method, we adopt a weighted-mean algorithm with a $1\sigma$ noise defined in each survey field. We apply $3\sigma$ clipping to remove shot noise and accidental false signals. Because our $z\ge3.1$ LAEs are found in the single field of SXDS, we simply obtain mean-combined images with no weighting for our $z\ge3.1$ LAEs. 
To make the median-combined image, we normalize our LAE images with the total fluxes, and perform median stacking
with a weight based on a signal-to-noise ratio of survey field.
Note that additional errors from the total flux measurements are included in this normalization process, 
and that the median-combined images have uncertainties lager than the mean-combined images
\footnote{
By this reason, the results from the mean-combined images are thought to be more reliable 
than those from the median-combined images (Section \ref{scale_lejgths_of_our_lahs_and_comparisons}).
}.
After stacking the images,
we degrade all of the stacked images to the same pixel size of $0\farcs$3
to construct a matched pixel size dataset of composite images
\footnote{
The PSF size of all images is matched to $1\farcs32$ in FWHM by the image smoothing described in (i).
}
. 
The composite images are presented in Figures \ref{fig:image_mean} and \ref{fig:image_median},
and the radial SB profiles of our LAEs in the Ly$\alpha$ images are shown in Figure \ref{fig:radi_allz}.
In Figures \ref{fig:image_mean} and \ref{fig:image_median},
the sources in the Ly$\alpha$ images, especially for $z=6.6$ LAEs, 
appear slightly elongated. Because we display the images with color scales down 
to $\sim 1\sigma$ noise levels, the elongated features are probably made
by noise near the outskirts of sources. 
\end{description}

We stack images of point sources to obtain radial SB profiles of PSFs.
These PSFs are used to investigate a radial profile of PSF near the source center,
and we hereafter refer to these PSFs as small-scale PSFs. 
Similarly, we make composite images of 100 saturated point sources to investigate extended tails of PSFs, namely large-scale PSFs,
following the procedure of \citet{feld13}.
The stacking of point sources is performed in the same manner as LAEs. 
In Section \ref{systematic_errors}, we compare the radial profiles of PSFs 
with those of LAEs, evaluating systematic uncertainties of PSFs. 

We randomly select sky regions that include no objects within a $45\arcsec\times45\arcsec$ area, 
and define the images in the sky regions as sky images. 
We obtain sky images whose numbers are the same as the Ly$\alpha$ or continuum images of LAE samples, 
and stack the sky images to produce a composite sky image in the same manner as LAE images.
We repeat to make a composite sky image for one thousand times.
In this way, we make one thousand composite images from the sky images.
Using these one thousand composite sky images, we estimate uncertainties of LAE profiles.
We make a histogram of composite sky-image fluxes measured in an area same as that given for our LAE profile estimates,
and confirm that the histogram has a nearly Gaussian distribution. We define the standard deviation of the
distribution as one sigma error. Table \ref{tab:z} presents SB limits obtained by this procedure.
To evaluate the systematic errors given by spatially-correlated noise such discussed in
Section 5.2 of \citet{gawiser06a}, we measure errors for various sky areas, $A_{\rm sky}$, 
with our composite sky images, and investigate the relation between the error values and $A_{\rm sky}$. We find that 
the errors are not explained by the simple Poisson statistics, i.e. $A_{\rm sky}^{0.5}$, 
but by $A_{\rm sky}^{0.6}$ that indicates the existence of systematic errors, 
which is similar to the results for the broadband images of \citet{gawiser06a}.
To include these systematic errors of spatially-correlated noise into our $1\sigma$ uncertainties,
we do not scale a $1\sigma$ error of one specific area by $A_{\rm sky}^{0.5}$,
but obtain errors of the sky-image areas (+shape) exactly the same as those of 
apertures used for our radial profile estimates in the following sections.
Note that the SB limits given in Table \ref{tab:z} are those measured in
an area of 1 arcsec$^2$ with no scaling of noise by the size of area.

Figure \ref{fig:radi_allz} represents the profiles of the small-scale PSFs and compares these profiles with
those of LAEs. Continuum profiles of all LAE samples roughly follow the small-scale PSFs, while 
Ly$\alpha$ profiles appear to be more extended than the small-scale PSFs. Note that the extended Ly$\alpha$ emission 
in our $z=3.7$ LAE sample is marginally detected, due to its small sample size and the shallow $NB570$ data.
In the two stacking methods, we find extended Ly$\alpha$ emission beyond the small-scale PSFs.
The remaining question is whether systematics including the large-scale PSFs mimic
the extended Ly$\alpha$ profiles.

\begin{figure}
	\begin{center}
		\includegraphics[scale=0.45]{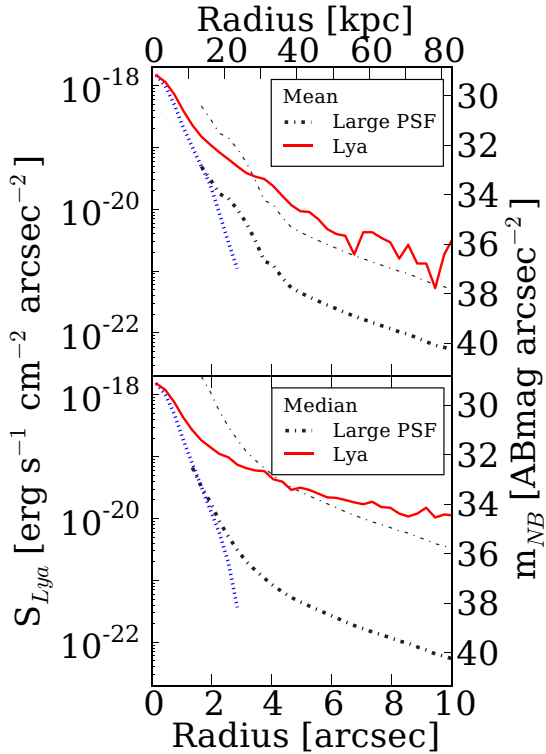} 
	\end{center}
	\caption{
	Radial SB profiles of LAEs and PSFs in the Ly$\alpha$ images of $NB387$.
	The red-solid lines represent the Ly$\alpha$ profiles of LAEs.
	The 
	blue
	and black dashed lines denote the small and large-scale PSFs, respectively.
	The gray lines are the large-scale PSF profiles with offsets in SB for the shape comparison with the Ly$\alpha$ profiles.
	Top and bottom panels show the results of mean and median-combined methods, respectively.
	}
	\label{fig:lPSF}
\end{figure}

\begin{figure} 
	\begin{center}
		\includegraphics[scale=0.26]{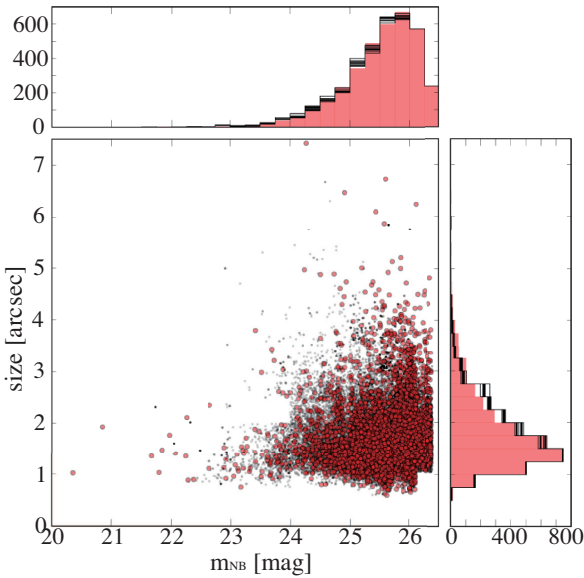}
	\end{center}
	\caption{
	Size and NB-magnitude distribution of LAEs and non-LAEs of $NB387$ images.
	In the central panel, the red and black circles represent LAEs and non-LAEs, 
	respectively. Similarly,
	in the top and right panels, red and black histograms denote LAEs and non-LAEs, respectively.
	}
	\label{fig:sample_nonlae}
\end{figure}

\section{Systematic Errors}
\label{systematic_errors}
It is argued that systematic uncertainties of the image stacking can produce a spurious extended profile of Ly$\alpha$ in composite images. \citet{feld13} have claimed that there are 
two 
systematic sources that produce a spurious extended Ly$\alpha$ profile. 
One 
is the large-scale PSF that could be made by instrumental and atmospheric effects. The 
other 
is systematic errors of flat-fielding. 
In addition to these 
two 
sources of systematics, we think that residuals of sky subtraction may also mimic extended Ly$\alpha$ profiles. Here, we examine the impacts on these systematic uncertainties in two ways.

\subsection{Large-Scale PSF Errors}
\label{large-scale_psf_errors}
Figure \ref{fig:lPSF} compares the small and large-scale PSF profiles with the Ly$\alpha$ profiles of LAEs in
the $NB387$.
There are spatially extended Ly$\alpha$ profiles of LAEs in Figure \ref{fig:lPSF}, but
here we investigate whether these spatially extended Ly$\alpha$ profiles are real or spurious signals. 
Because the central profiles of large-scale PSFs are contaminated by saturation, we connect the large-scale PSF to the small-scale PSF (Section \ref{image_stacking}) in the radius range with no saturation effects. Figure \ref{fig:lPSF} indicates that the large-scale PSFs provide fluxes much fainter than the Ly$\alpha$ emission by $\gtrsim 2-3$ magnitudes, and that the profile shape of large-scale PSF are clearly different from those of extended Ly$\alpha$. 
We thus confirm that the large-scale PSFs do not mimic the extended Ly$\alpha$ profile of our LAEs.

\begin{figure*} 
	\begin{center}
		\includegraphics[scale=0.25]{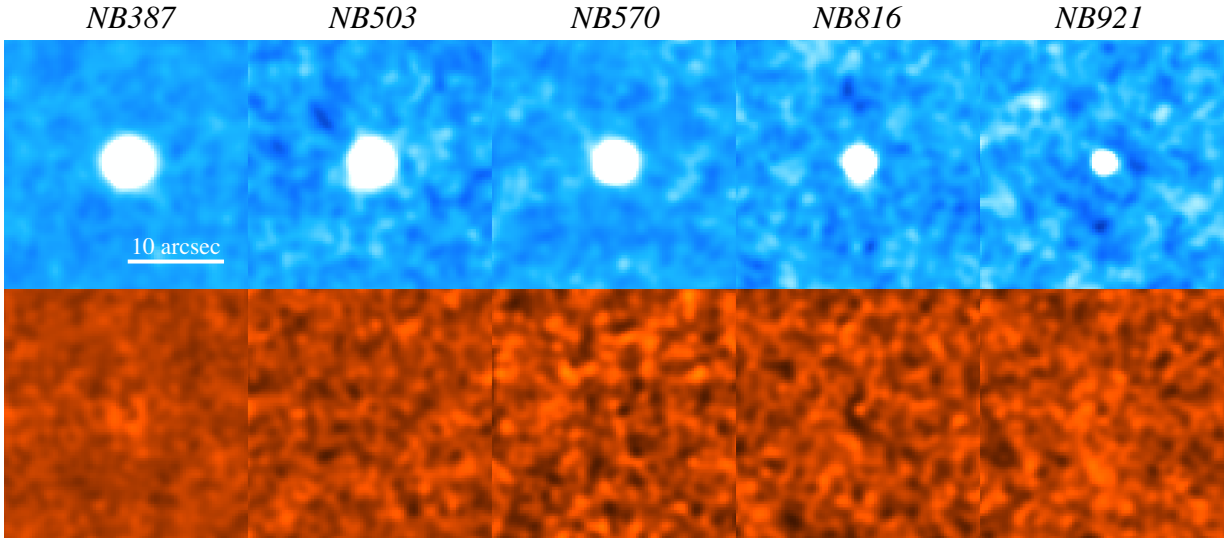}
	\end{center}
	\caption{
	Composite continuum (top panels) and Ly$\alpha$ (bottom panels) images of non-LAEs
	that are made by the mean-combined method.
	From left to right panels, we show $NB387$, $NB503$, $NB570$, $NB816$, and $NB921$ images.
	}
	\label{fig:image_nonlae_mean}
\end{figure*}

\begin{figure*} 
	\begin{center}
		\includegraphics[scale=0.25]{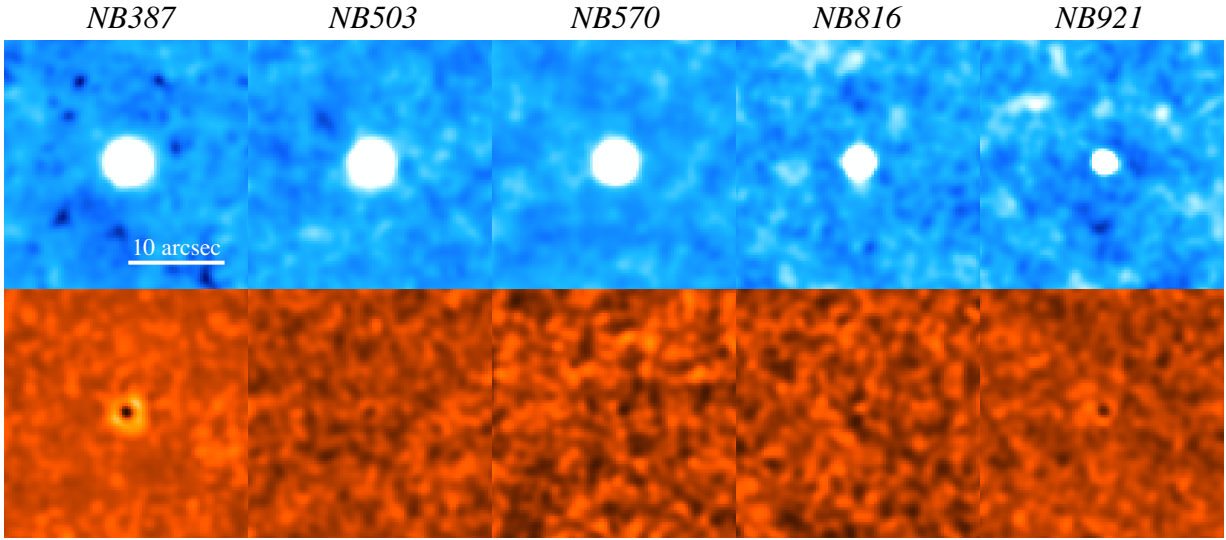}
	\end{center}
	\caption{
	Same as Figure \ref{fig:image_nonlae_mean}, but for the median-combined method.
	}
	\label{fig:image_nonlae_medi}
\end{figure*}

\subsection{Tests for All Systematic Errors}
\label{tests_for_all_systematic_errors}
In Section \ref{large-scale_psf_errors}, we rule out the possibility that the large-scale PSFs give spurious signals mimicking extended Ly$\alpha$. However, there are a number of unknown systematics that include flat-fielding and sky-subtraction errors. Although the large-scale flat-fielding error may not be a major source of systematics in our high-quality images of Suprime-Cam, one needs to carefully evaluate total errors contributed from all sources of systematics. We carry out image stacking for objects that are not LAEs, which are referred to as non-LAEs. Because non-LAEs have no intrinsically extended emission-line halos like LAHs, extended profiles of non-LAE composite images should be given by a total of all systematic effects. We thus make composite images of non-LAEs, and investigate how much systematics the total of all systematic errors produce.

\begin{figure*} 
	\begin{center}
		\includegraphics[scale=0.35]{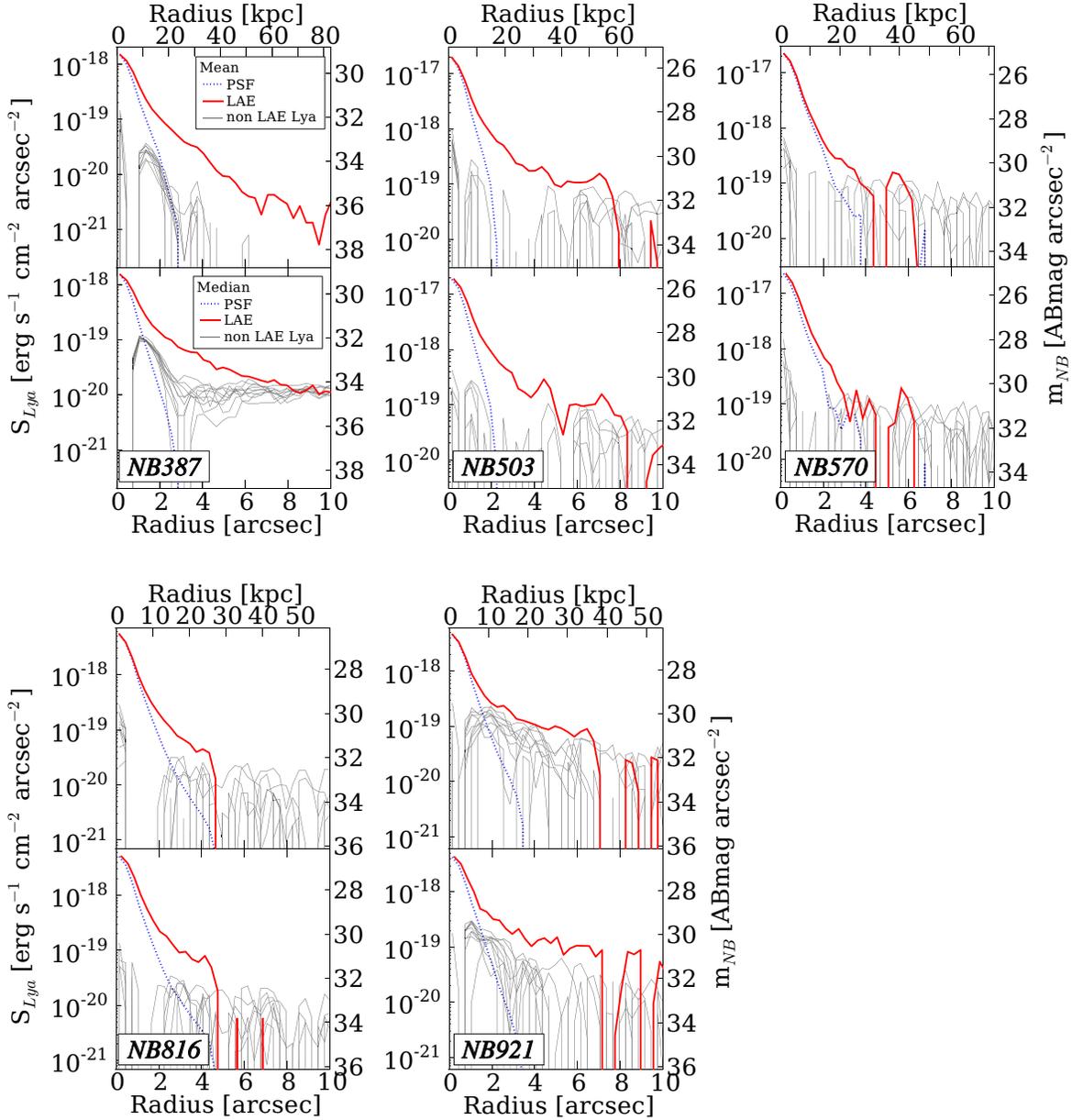}
	\end{center}
	\caption{
	Radial profiles of LAEs, non-LAEs, and PSFs in the Ly$\alpha$ images of 
$NB387$, $NB503$, $NB570$, $NB816$, and $NB921$.
	The red and black lines represent LAEs and non-LAEs of ten realizations, respectively.
	The blue
	dotted lines denote the small-scale PSFs.
	Top and bottom panels show the results of mean and median-combined methods, respectively.
	}
	\label{fig:radi_nonlae}
\end{figure*}

First, we randomly choose non-LAEs with the same number as our LAEs. These non-LAEs have size and NB-magnitude distributions same as those of our LAE samples (Figure \ref{fig:sample_nonlae}). To make a Ly$\alpha$ image of the non-LAE sample, we normalize a composite continuum image to match the total flux of a composite NB image, and then subtract the continuum image from the composite NB image. We investigate whether an artificial extended profile appears in the Ly$\alpha$ image of non-LAEs.
To reveal uncertainties of this estimate, we repeat it ten times for ten realizations.
Figures \ref{fig:image_nonlae_mean} and \ref{fig:image_nonlae_medi} present composite images of non-LAEs given by the mean and median-combined methods, respectively. We identify no significant extended profiles in these images. We should note that a ring-like structure found at the source centers of $NB387$ data is attributable to the slight differences of small-scale PSF profiles of continuum and NB images, which are irrelevant to the extended profiles. 
Radial profiles of Ly$\alpha$ images of non-LAEs are shown with black lines in Figure \ref{fig:radi_nonlae}. 
In the $NB387$ panels of Figure \ref{fig:radi_nonlae},
we find that there are artificial extended profiles at the level of $\lesssim 10^{20}$ erg s$^{-1}$ cm$^{-2}$ arcsec$^{-2}$,
but that these profiles of artifacts are significantly different from those of LAEs in the SB-profile amplitude and shape. 
These tests confirm that the total of systematic uncertainties do not produce spatially extended profiles similar
to the Ly$\alpha$ profiles of LAEs in the $NB387$ images.
Based on these tests, we conclude that the spatially extended profiles of $NB387$ LAEs 
in our Ly$\alpha$ images are real, and regard these spatially extended Ly$\alpha$ features as LAHs.
Similarly, the panels of $NB503$ and $NB816$ show that the differences of profiles between LAEs and artifacts exist clearly,
and the LAHs of $NB503$ and $NB816$ are also identified.
The profile of median-combined image of $NB921$ is different from that of artifacts in the SB-profile amplitude,
but the profile of mean-combined image of $NB921$ is only beyond that of the artifacts with a small offset.
These results indicate that there exist LAHs of $NB921$ LAEs, although
we need to carefully discuss the $NB921$ results below.
In Figure \ref{fig:radi_nonlae}, the extended profiles of $NB570$ are indistinguishable from those of artifacts.
Note that the $NB570$ data are made of a small number of observation image frames, and that
the quality of $NB570$ data is not as good as the other narrowband images \citep{ouchi08}.
Due to the poor quality of $NB570$ data,
significant signals of extended $NB570$ profiles beyond the artifacts are probably not identified.
We find that the extended profiles of $NB570$ ($z=3.7$) are artifacts, and
do not discuss the profiles of $z=3.7$ LAEs in the following sections.

\section{Results}
\label{results}
In Sections \ref{data_and_analysis} and \ref{systematic_errors},
we have identified LAHs in our Ly$\alpha$ images, 
and confirmed that the LAH signals are not produced by systematic errors. 
In this section, we estimate the scale lengths of our LAHs
based on the radial profiles of composite Ly$\alpha$ images.

\subsection{Definition of the Scale Length}
Following the previous studies \citep{stei11,matsu12,feld13},
we define the scale length of $r_n$ with the exponential profile of
\begin{equation}
	S(r) = C_n \exp{(-r/r_n)},
\end{equation}
where $S(r)$, $r$, and $C_n$ are the SB of radial profile, radius, and normalization factor, respectively. We carry out the profile fitting in a radius range from $r=2\arcsec$ to $40$ kpc
for all of our LAE samples. This radius range allows us to obtain $r_n$ with negligible contaminations of PSF
\footnote{
The FWHM PSF size is $1\farcs32$ given by the image smoothing in Section \ref{image_stacking}.
The smoothing makes the data points strongly correlated each other within the scale of $1\farcs32$.
}
 ($r>2\arcsec$) 
and 
reasonably high statistical accuracies ($r<40$ kpc).

\subsection{Scale Lengths of our LAHs and Comparisons}
\label{scale_lejgths_of_our_lahs_and_comparisons}
We perform the profile fitting to our LAHs at $z=2.2$, $3.1$, $5.7$, and $6.6$, 
and estimate $r_n$ as well as $C_n$ values. Note that the profile fitting
results are not presented for our sample of $z=3.7$ (i.e. $NB570$) whose
extended profiles are made of artifacts (Section \ref{tests_for_all_systematic_errors}).
All of the LAH-detected samples,
except the one of $z=2.2$, have
similar Ly$\alpha$ luminosity limits of $L_{\rm Ly\alpha}=1-3\times10^{42}$ erg s$^{-1}$.
Since our
LAE sample 
of $z=2.2$
reaches a Ly$\alpha$ luminosity limit fainter than those of our $z\ge 3.1$ samples,
we make a subsample of $z=2.2$ LAEs with 
a Ly$\alpha$ luminosity down to $1\times 10^{42.0}$ erg s$^{-1}$
that consists of 2115 LAEs. 
\footnote{
Our profile fitting is carried out
both for the entire sample and subsample of $z=2.2$ LAEs.
}
In this way, we obtain samples of LAEs at $z=2.2-6.6$ with an average Ly$\alpha$ luminosity limit of
$L_{\rm Ly\alpha}\gtrsim 2\times 10^{42.0}$ erg s$^{-1}$.
Table \ref{tab:z} summarizes our best-fit parameters and those in the literature.
Below, we show details of our results, and compare our results with
those from previous studies.

\paragraph*{ {\it z = 2.2} ---} 
We find that the best-fit scale lengths of our LAHs at $z=2.2$
are 
$r_n=10.0^{+0.38}_{-0.36}$ and $7.9^{+0.56}_{-0.49}$ kpc
($14.0^{+9.0}_{-3.9}$ and 11.1$^{+1.2}_{-0.97}$ kpc)
for the entire and sub-samples, respectively, which are
estimated by the mean-combined (median-combined) method.
The SB limits of our composite Ly$\alpha$ images reach depths of 
$1.3\times10^{-20}$ and $1.6\times10^{-20}$ erg s$^{-1}$ cm$^{-2}$ arcsec$^{-2}$
in our entire and sub-samples, respectively.
\citet{stei11} have detected LAHs by stacking NB images of $92$ LBGs with a mean spectroscopic redshift of $\langle z\rangle=2.65$. The scale lengths of the LAHs in \citet{stei11} are $r_n=25.2$ kpc (mean) and $17.5$ kpc (median) which are $\sim 2$ times larger than our values. 
This discrepancy is probably originated from the difference of galaxy populations, LBGs and LAEs, and detailed discussions
are given in Section \ref{scale-length_dependence_on_galaxy_population_and_envrinment}.
On the other hand, \citet{feld13} have found no extended Ly$\alpha$ emission in their composite image made from 187 LAEs at $z=2.06$. The difference of our and \citeauthor{feld13}'s results are probably due to their SB limit shallower than those of our composite images by $1-2$ orders of magnitudes.

\paragraph*{ {\it z = 3.1} ---} 
The best-fit scale lengths of our $z=3.1$ LAHs are
$r_n=9.3^{+0.48}_{-0.53}$ (mean) and $6.3^{+2.5}_{-1.4}$ kpc (median) down to the SB limit of 
$1.7\times10^{-19}$ erg s$^{-1}$ cm$^{-2}$ arcsec$^{-2}$. 
\citet{matsu12} have identified LAHs at $z=3.1$ with 
their large LAE samples, and found that their scale lengths depend on the surface number density of LAEs. The scale lengths obtained by \citet{matsu12} are $9.1$ and $20.4$ kpc in the LAE's lowest and highest-density environments, respectively.
The scale length in the lowest-density environment is comparable with our $z=3.1$ values within the fitting errors,
while the scale length in the highest-density environment is larger than our $z=3.1$ values.
We revisit this issue of environment in Section \ref{scale-length_dependence_on_galaxy_population_and_envrinment}.
\citet{feld13} have marginally detected LAHs by stacking $241$ and $179$ LAEs at $z=3.10$ and $3.21$, respectively. 
Their scale lengths are $r_n=2.8-8.4$ kpc, which are also comparable to our scale lengths.

\paragraph*{ {\it z $\geq$ 5.7} ---} 
For our $z=5.7$ and $6.6$ LAHs, we obtain
$r_n= 5.9^{+0.65}_{-0.53}$ and $12.6^{+3.3}_{-2.4}$ kpc
with the mean-combined data down to the SB limits of 
$5.5\times10^{-20}$ and $1.8\times10^{-19}$
erg s$^{-1}$ cm$^{-2}$ arcsec$^{-2}$, respectively.
Similarly, for the median-combined data,
the scale lengths of $z=5.7$ and $6.6$ LAHs are
7.7$^{+1.9}_{-1.3}$ and 13.9$^{+0.75}_{-0.99}$ kpc, respectively.
There are no previous results that can be compared with these scale lengths of our results. 
Note that \citet{jian13} have found no LAHs around $z=5.7$ and $6.6$ LAEs
with their small sample of 43 and 40 LAEs, respectively.
Although \citet{jian13} claim that their stacked images reach the SB limit of 
$1.2\times 10^{-19}$ erg s$^{-1}$ cm$^{-2}$ arcsec$^{-2}$ at the 1 sigma level
that is comparable to our SB limits within a factor of 
$\sim 2$
, the error estimates
are unclear in \citet{jian13}. The results of no-LAH detection of \citet{jian13} are 
probably due to their small statistics.

The results shown above indicate that the scale lengths from the median-combined images
are comparable with those from the mean-combined images within the 1 sigma uncertainties
for the most of measurements. We confirm that our results do not significantly depend on the
choice of statistics. 
Moreover, the results are not affected by the contamination of our LAE samples (Section \ref{data_and_analysis}),
since, under the influence of contamination,
the results of mean-combined images should be different 
from those of median-combined images.
Because the median-combined image results include additional
uncertainties explained in Section \ref{image_stacking}, we regard the mean-combined image
measurements as reliable results. We hereafter make discussions based on the mean-combined
image results, unless otherwise specified.

\begin{figure*} 
	\begin{center}
		\includegraphics[scale=0.4]{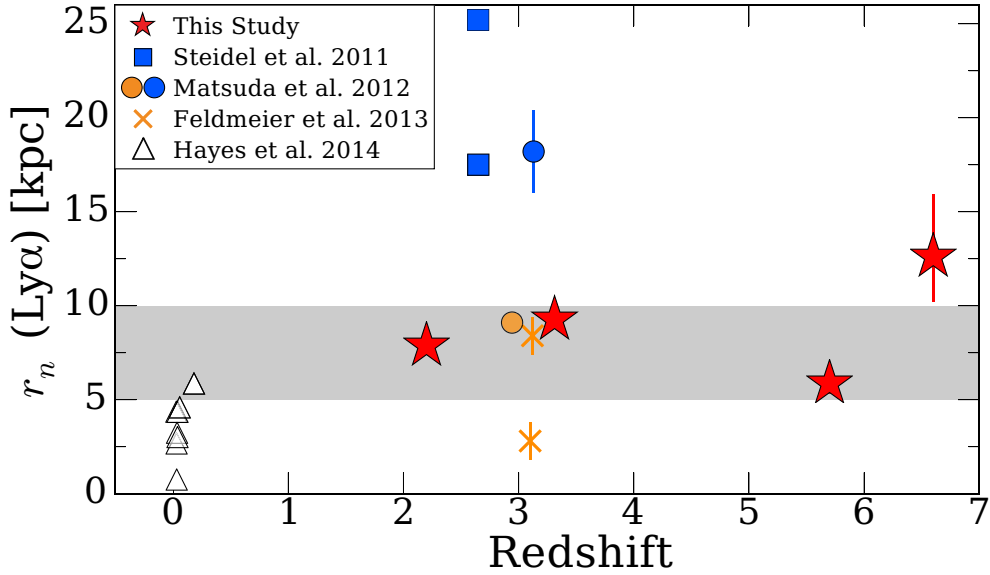}
	\end{center}
	\caption{
	Scale length as a function of redshift. 
	The red stars represent the scale lengths estimated by the mean-combined method in this study.
	The gray shade indicates the range of $5-10$ kpc in which our reliable scale length estimates at $z=2.2-5.7$ fall.
	The blue squares and circle denote scale lengths of LAHs found 
	around LBGs of \citet{stei11} and \citet{matsu12}, respectively.
	The orange circle represents the lowest-density region LAEs of \citet{matsu12}, which are slightly shifted along the abscissa for clarity. 
	The orange crosses are the scale lengths of LAHs around LAEs \citep{feld13}. 
	For reference, we plot Ly$\alpha$ Petrosian radii of local LAEs with the open triangles \citep{hay13a,hay13b}.
	Here, we regard galaxies with a Ly$\alpha$ equivalent width of $\geq 20$\AA\ as local LAEs in \citet{hay13a,hay13b}, 
	and plot Petrosian radii of their samples that meet this equivalent-width criterion.
	}
	\label{fig:size}
\end{figure*}

\subsection{Size Evolution of LAHs}
\label{size_evolution_of_lahs}
We investigate evolution of LAH scale lengths in the redshift range of $z=2.2-6.6$. Figure \ref{fig:size} presents the scale lengths as a function of redshift.
The scale lengths at $z=2.2-5.7$
fall in the range of $5-10$ kpc that includes measurement uncertainties (Table \ref{tab:z}). 
Thus, Figure \ref{fig:size} indicates no evolution of the scale lengths from $z=2.2$ to $5.7$.
In the redshift range of $z=5.7-6.6$, there is a hint of increase of scale length. The scale lengths in the mean-combined method increase from $z=5.7$ to $6.6$ over the fitting errors, $r_n=5.9^{+0.65}_{-0.53}$ kpc at $z=5.7$ and $r_n=12.6^{+3.3}_{-2.4}$ kpc at $z=6.6$.
Because, in Section \ref{tests_for_all_systematic_errors}, we find that the $z=6.6$ LAH profile of mean-combined image 
would be influenced by artifacts, this increase may not be real based on the data of mean-combined images. 
However, the same trend is also found in the scale lengths obtained by the median-combined method
with no signature of artifacts; 7.7$^{+1.9}_{-1.3}$ and 13.9$^{+0.75}_{-0.99}$ kpc at $z=5.7$ and $6.6$, respectively.
Thus, our data indicate the increase of scale length from $z=5.7$ and $6.6$, 
although the significance of increase is only beyond the statistical error.

\section{Discussions}
\label{discussions}
\subsection{Do LAHs Really Exist?}
The existence of LAHs around high-redshift galaxies is under debate 
\citep{feld13,jian13}. In Figure \ref{fig:radi_allz}, we identify statistically-significant extended Ly$\alpha$ emission around our LAEs at $z=2.2-6.6$ based on our unprecedentedly large LAE samples that allow us to achieve $\sim 10-100$ times deeper SB limits than those of typical previous studies (e.g., \citealt{stei11,feld13}; see Table \ref{tab:z}). In Section \ref{systematic_errors}, we examine potential systematic errors that would mimic LAHs, and find that the large-scale PSF of instrumental and atmospheric effects cannot produce radial profiles of our extended Ly$\alpha$ emission in SB and shape (Figure \ref{fig:lPSF}). Besides the large-scale PSF, there are a number of potential systematic effects, such as flat-fielding and sky subtraction errors (e.g., \citealt{feld13}) as well as unknown systematics. To reveal the total systematic errors involved in our data and analysis, we stack non-LAEs (Figure \ref{fig:sample_nonlae}) in the same manner as our LAEs, and carry out the empirical tests. We find that there exist systematic errors that make an extended emission signal, but that no systematic errors can make a radial
profile with the SB amplitude and shape similar to those of our extended Ly$\alpha$ emission of 
LAEs, except for our sample of $z=3.7$ LAEs
whose data quality is poor
(Figure \ref{fig:radi_nonlae}). 
Thus, we conclude that we definitively identify LAHs around the LAEs with our data by our analysis technique.
Our results indicate that LAEs commonly possess LAHs 
at $z=2.2-6.6$.

\subsection{Physical Origin of LAHs}
\label{physical_origin_of_lahs}
In theoretical studies, LAHs are thought to be produced primarily by two physical mechanisms: 1) the resonant scattering of Ly$\alpha$ in the CGM and/or IGM (e.g., \citealp{lau07,zhen11,dijk12,verha12,jee12}), and 2) the cold streams (e.g., \citealp{fau10,goe10,ros12}). 
Our observational results imply that LAHs would be made by the former mechanism.

First, we discuss the possibility of resonant scattering.
Ly$\alpha$ photons escaping from a galaxy are scattered by the neutral hydrogen in the CGM within a few 100 kpc from the center of galaxy, making the Ly$\alpha$ emission extended more than that of stellar continuum. Theoretical predictions indicate that the SB profiles of LAHs are determined by a combination of the ISM dynamics and distribution (e.g., \citealp{lau07,zhen11,dijk12,verha12}). \citet{lau09a,lau09b} have carried out Monte Carlo radiative transfer simulations of Ly$\alpha$ propagating through the ISM with various kinematic properties, and found that the extent of LAHs is $r\sim50-100$ kpc. \citet{verha12} have predicted that the clumpy and inhomogeneous ISM produces an LAH with a characteristic radius of $r\sim10-20$ kpc in their  cosmological simulations. The LAHs from our observations also extend up to a radius of $r\sim30-80$ kpc
similar to those of the simulations (Figure \ref{fig:radi_allz}). 
Moreover, similar LAHs are predicted for the neutral IGM scattering Ly$\alpha$ photons at the epoch of reionization  (e.g., \citealp{zhen11,jee12}).
Thus, the sizes of observed LAHs are comparable to those of Ly$\alpha$ scattering models, which indicate that
the major physical origin of the LAH is probably the resonant scattering of Ly$\alpha$ photons in neutral hydrogen of the CGM and/or IGM.

Second, we examine whether the cold streams would be a major mechanism of the LAH formation. Cosmological hydrodynamic simulations have predicted that high-redshift ($z\geq2$) galaxies assemble baryon via accretion of relatively dense and cold gas ($\sim10^4$ K) that represents the cold streams (e.g., \citealp{keres05,dek09a,dek09b}). At the temperature of $\sim10^4$ K, the cold gas could primarily emit Ly$\alpha$ (e.g., \citealp{far01}), which would produce an LAH around galaxies  \citep{fau10,goe10,ros12}. Cosmological simulations have predicted that Ly$\alpha$ emission are more extended for a higher dark halo mass of host galaxies. \citet{ros12} have found that cold streams in dark halos with a mass of $\geq10^{12}$ M$_\odot$ produce extended Ly$\alpha$ structures, but that Ly$\alpha$ emission is centrally-concentrated 
($< 20$ kpc in diameter) in less-massive dark halos with a mass of $\sim10^{11}$ M$_\odot$. Because the typical dark halo mass of LAEs is estimated to be $\sim10^{11\pm1}$ M$_\odot$ (\citealp{ouchi10} and reference therein), the cold streams would not make an LAH as large as
$30-80$ kpc found in our data.
The cold streams are probably not the major mechanism of the LAH formation.

\subsection{Scale-Length Dependence on Galaxy Population and Environment}
\label{scale-length_dependence_on_galaxy_population_and_envrinment}
In Section \ref{scale_lejgths_of_our_lahs_and_comparisons}, we find differences of scale lengths between
measurements of this and some previous studies. 

First, at $z\sim 2$, the scale lengths from this study 
are smaller than those from \citet{stei11} by a factor of $\sim 2$ (Figure \ref{fig:size}). Note that the scale lengths of \citet{stei11} 
are estimated for their LBGs, while our study obtains the scale lengths of LAEs.
By definition, Ly$\alpha$ emission of LAEs are brighter than that of LBGs, at a given SFR or UV continuum on average. 
Theoretical studies suggest that the resonance line of Ly$\alpha$ can escape from galaxies with 
a low column density of neutral hydrogen in the CGM \citep{verha12,zhen11}.
This physical picture of Ly$\alpha$ escape is confirmed by recent imaging and spectroscopic observations \citep{shib14a,shib14b}.
Based on this picture, Ly$\alpha$ photons produced in star-forming regions of LAEs reach the observer
with little resonant scattering. The observer thus finds that Ly$\alpha$ profiles of LAEs are more centrally concentrated 
than those of LBGs, and obtains a bright Ly$\alpha$ luminosity in
the central core of Ly$\alpha$ profile of LAEs above the shallow detection limit of non-stacked images.
On the other hand, LBGs have more resonant scattering in the CGM than LAEs, and Ly$\alpha$ profiles of LBGs
are largely extended. This is consistent with the fact that
the SB of LBGs of \citealt{stei11} ($\sim 10^{-18}$ erg s$^{-1}$ cm$^{-2}$ Hz$^{-1}$ arcsec$^{-2}$ at 30 kpc)  is 
one-order of magnitude brighter than that of our LAEs ($\sim 10^{-19}$ erg s$^{-1}$ cm$^{-2}$ Hz$^{-1}$ arcsec$^{-2}$ at 30 kpc).
The scale-length difference of our LAEs and \citeauthor{stei11}'s LBGs indicates that 
LAEs do not have much neutral hydrogen CGM that scatters Ly$\alpha$, and 
that the Ly$\alpha$ profiles of LAEs are steeper than LBGs.

Second, at $z \sim 3$, the LAH scale lengths of our LAEs are comparable with
those of the lowest-density environment LAEs of \citet{matsu12}, but 
lower than the highest-density environment LAEs of \citet{matsu12} (Figure \ref{fig:size}).
Here, the galaxy populations of our and \citeauthor{matsu12}'s samples are the same. 
The difference of scale lengths is probably explained by the environment.
Clustering analysis with our LAE samples indicates that our survey field at $z=3.1$ 
is not a high density region \citep{ouchi08,ouchi10}. Our results confirm that
low-density environment LAEs have a moderately small scale length
of $r_n\simeq 5-10$ kpc, and support the idea of the environmental effect on LAHs 
that is claimed by \citet{matsu12}.
Theoretical studies have predicted that the SB profiles are affected by the galaxy environment. \citet{zhen11} have found the SB profiles of their LAHs at $z=5.7$ have three notable features separated at two radial positions; a central cusp at $r\leq0.2$ Mpc (co-moving), a relatively-flat part at $r\sim0.2-1$ Mpc, and an outer steep region at $r\geq1$ Mpc. The two radial positions of $r=0.2$ and $1$ Mpc are characterized by the one- and two-halo terms of dark matter halos. However, we do not find such features in the radial profiles of our LAHs. This is probably because our LAHs at $z=5.7$ are only found within an inner region of $r\leq 0.2$ Mpc, and/or the clustering strength of the SXDS field is weaker than that of Zheng et al. (2011; private communication). 
One needs deeper images than those of this study to investigate the environmental effect at the moderately high redshift of $z=5.7$.

\subsection{Size Evolution of LAHs}
We find no size evolution of LAHs ($r_n\simeq 5-10$ kpc) from $z=2.2$ to $5.7$ as we describe in Section \ref{size_evolution_of_lahs}. \citet{mal12} have measured the half-light radii of stellar distribution of LAEs, $r_c$, in the rest-frame UV, and found no size evolution ($r_c\sim1$ kpc on average) in the redshift range of 
$z\simeq 2-6$. These two results indicate that the size ratio of LAHs to stellar component is almost constant, $r_n/r_c \sim 5-10$, between $z=2.2$ and $5.7$. This ratio is comparable with those of $z\sim 2$ LBGs ($r_n/r_c \sim5-10$; \citealt{stei11}) and $z\sim 0-3$ LAEs ($r_n/r_c \sim2-4$; \citealt{matsu12,hay13a}). The former LBG results indicate that 
there is a scaling relation between the LAH and stellar-distribution sizes over the samples of LBGs and LAEs.
The latter should be comparable, because these results include a low-density environment LAE sample similar to ours.
This no evolution of $r_n$ and $r_n/r_c$ at $z=2.2-5.7$ is interesting, because
similar trends of no evolution of LAE's physical properties are found in this redshift range.
Spectral energy distribution (SED) fitting of our LAE samples have revealed that stellar population
of our LAEs do not evolve significantly in the redshift range of $z=2.2-5.7$ (\citealt{ono10a,ono10b,naka12}, see also \citealp{gaw06,pen07,pen10,fin11,mal12}). 
Moreover, hosting dark halos of LAEs are typically 
$M_{\rm DH}=10^{11\pm 1}$ M$_\odot$ at $z=2-7$ (\citealp{ouchi10} and reference therein), and do not evolve.
These observational results of no evolution of $r_n$, $r_n/r_c$, stellar population, and dark halo mass would indicate
that LAEs are the population in a specific stage of galaxy evolution, which NB observations can snapshot.
 
We find a possible increase of scale length from $z=5.7$ to $6.6$ in Section \ref{size_evolution_of_lahs} (see Figure \ref{fig:size}). 
Again, clustering analysis with our LAEs indicates that our survey field at $z=6.6$ is not a high density region \citep{ouchi10},
and it is unlikely that this increase is due to the environmental effect.
Because there is no significant increase of scale lengths in the redshift range of $z=2.2-5.7$, this sudden increase
from $z=5.7$ to $6.6$ may be explained by cosmic reionization.
Signatures of the increase in the neutral hydrogen fraction of IGM ($x_{\text{HI}}$) 
at $z\geq7$ have been 
found by many observational studies based on Ly$\alpha$ luminosity functions of LAEs or 
the Gunn-Perterson test of QSOs (e.g., \citealp{fan06,ota08,ouchi10,kashik11,goto11,shib12}). 
Numerical simulations show that Ly$\alpha$ emission around star-forming galaxies 
is extended due to the neutral hydrogen in the IGM (e.g., \citealp{zhen11,jee12}). 
\citet{jee12} have predicted that the SB profile of LAHs becomes flatter in the IGM with 
a high $x_{\text{HI}}$.
The relatively large scale length of our LAHs at $z=6.6$ may come 
from a more neutral IGM at the epoch of reionization,
although 
the reliability of increase is not very high in statistics (Figure \ref{fig:size}) and systematics (Section \ref{tests_for_all_systematic_errors};
see Section \ref{size_evolution_of_lahs} for the details).
To conclude this trend, one needs a large amount of high-quality NB data such obtained by
the upcoming survey of Subaru Hyper Suprime-Cam (HSC).

\section{Summary}
\label{summary}
We investigate diffuse LAHs of LAEs with composite Subaru NB images of ($3556$, $316$, $100$, $397$, $119$) LAEs at $z=$ ($2.2$, $3.1$, $3.7$, $5.7$, $6.6$), and discuss their physical origin and size evolution. The major results of our study are summarized below.
\begin{enumerate}
	\setlength{\leftskip}{0.5cm}
	\renewcommand{\theenumi}{\arabic{enumi}.}
	\item We detect extended Ly$\alpha$ emission around LAEs at 
	$z=2.2-6.6$
	in the composite images. We carefully examine 
	whether the radial profiles of Ly$\alpha$ emission can be made by systematic errors that include a large-scale PSF of 
	instrumental and atmospheric effects.
	Stacking Ly$\alpha$ images of randomly-selected non-LAEs, we confirm that the combination of all systematic errors cannot produce the 
	extended Ly$\alpha$ emission in our composite images
	of all redshifts of $z=2.2-6.6$, except for $z=3.7$ whose data have a quality poorer than the others.
	We thus conclude that the extended Ly$\alpha$ emission found in our composite images
	are real, and regard the extended Ly$\alpha$ emission as LAHs.
	\item We investigate radial SB profiles of our LAHs, and measure their characteristic exponential scale lengths. The scale lengths are estimated to be 
	$\simeq 5-10$ kpc at $z=3.1-5.7$ and 12.6$^{+3.3}_{-2.4}$ kpc at $z=6.6$ for LAE samples with 
	$L_{\rm Ly\alpha}\gtrsim 2\times 10^{42}$ erg s$^{-1}$
	(Figure \ref{fig:size} and Table \ref{tab:z})
	.
	The comparison with 
	the LAH scale lengths given by previous LBG and LAE studies would indicate that the radial profiles of LAHs depend on galaxy populations 
	and environment:
	LAEs have a centrally-concentrated Ly$\alpha$ profile, and LAEs in low-density regions possess less extended LAHs.
	\item We identify no evolution of scale lengths ($5-10$ kpc) from $z=2.2$ to $5.7$ beyond our measurement uncertainties.
	Combining with no size evolution of 
	LAHs and UV continuum emission of LAEs found by \citet{mal12} ($r_c\sim1$ kpc) over $z\sim2-6$, 
	we suggest the ratio of $r_n/r_c$ 
	to be nearly constant ($r_n/r_c \sim 5-10$) over the redshift range, $z=2.2-5.7$. This no evolution of $r_n/r_c$ probably indicates that the 
	population of LAEs at $z=2.2-5.7$ would be in the same evolutionary stage.
	\item We find a possible increase in the scale length from $z=5.7$ to $6.6$. This sudden increase only found at $z>6$ 
	may be a signature of increasing neutral hydrogen of IGM that scatters 
	Ly$\alpha$ photons due to the cosmic reionization. This finding would support the theoretical model predictions \citep{jee12},
	although there remain the possible problems of statistical and systematic effects on the measurements of $r_n$ at $z=6.6$.
	The upcoming surveys such with Subaru HSC will allow us to test whether this hint of increase is real or not.
\end{enumerate}

\section*{Acknowledgements}

We thank Alex Hagen, Caryl Gronwall, Lucia Guaita, 
Yuichi Matsuda,
Michael Rauch, James Rhoads, Anne Verhamme and Zheng Zheng
for useful comments and discussions.
This work was supported by World Premier International Research
Center Initiative (WPI Initiative), MEXT, Japan,
and KAKENHI (23244025) Grant-in-Aid for Scientific Research
(A) through Japan Society for the Promotion of Science
(JSPS). K.N. and S.Y. acknowledge the JSPS Research Fellowship
for Young Scientists.

{}


\end{document}